\documentclass[aps,prd,preprint,floats,superscriptaddress]{revtex4}
\usepackage{latexsym}

\usepackage{graphicx}

\def\bwt{\begin{widetext}}
\def\ewt{\end{widetext}}
\def\be{\begin{equation}}
\def\ee{\end{equation}}
\def\bea{\begin{eqnarray}}
\def\eea{\end{eqnarray}}
\def\bean{\begin{eqnarray*}}
\def\eean{\end{eqnarray*}}
\def\bary{\begin{array}}
\def\eary{\end{array}}
\def\bi{\bibitem}
\def\bit{\begin{itemize}}
\def\eit{\end{itemize}}

\def\ol{\overline}

\def\ub{\ol{u}}
\def\db{\ol{d}}
\def\sb{\ol{s}}

\def \b{{\cal B}}
\def \ca{{\cal A}}
\def \ko{K^0}
\def \ok{\overline{K}^0}
\def \s{\sqrt{2}}
\def \st{\sqrt{3}}
\def \sx{\sqrt{6}}

\begin{document}

\preprint{ANL-HEP-PR-02-073, EFI-02-56, hep-ph/0209272}
\bigskip

\title{Two-Body Cabibbo-Suppressed Charmed Meson Decays}

\author{Cheng-Wei Chiang}
\email[e-mail: ]{chengwei@hep.uchicago.edu}
\affiliation{HEP Division, Argonne National Laboratory
9700 S. Cass Avenue, Argonne, IL 60439}
\affiliation{Enrico Fermi Institute and Department of Physics,
University of Chicago, 5640 S. Ellis Avenue, Chicago, IL 60637}
\author{Zumin Luo}
\email[e-mail: ]{zuminluo@midway.uchicago.edu}
\affiliation{Enrico Fermi Institute and Department of Physics, 
University of Chicago, 5640 S. Ellis Avenue, Chicago, IL 60637}
\author{Jonathan L.~Rosner}
\email[e-mail: ]{rosner@hep.uchicago.edu}
\affiliation{Enrico Fermi Institute and Department of Physics, 
University of Chicago, 5640 S. Ellis Avenue, Chicago, IL 60637}

\date{\today}

\begin{abstract}
The singly-Cabibbo-suppressed decays of charmed particles governed by the quark
subprocesses $c \to s u \bar s$ and $c \to d u \bar d$ are analyzed using a
flavor-topology approach, based on a previous analysis of the Cabibbo-favored
decays governed by $c \to s u \bar d$.  Decays to $PP$ and $PV$, where $P$ is a
pseudoscalar meson and $V$ is a vector meson, are considered.  We include
processes in which $\eta$ and $\eta\,'$ are produced.
\end{abstract}


\maketitle


\section{INTRODUCTION}

The decays of charmed particles can provide useful information about
strong interactions.  The magnitudes and phases of weak couplings
governing these decays are well-specified in the standard electroweak
theory, so that decay amplitudes serve mainly to illuminate the relative
importance of various flavor topologies and their relative strong phases.
This has been shown both for Cabibbo-favored decays \cite{JR99} and
more recently for doubly-Cabibbo-suppressed processes \cite{CR02}.  In the
present article we extend these analyses to singly-Cabibbo-suppressed
processes.

The analysis of charmed particle decays has shown that flavor SU(3) symmetry
is qualitatively obeyed, but important symmetry-breaking effects can be
identified \cite{hist}.  As one example, an argument using the U-spin
subgroup of SU(3) predicts the rates for $D^0 \to \pi^+ \pi^-$ and $D^0 \to K^+
K^-$ to be equal, but they differ by a factor of about three.  This effect
can be understood on the basis of SU(3)-breaking in decay constants and
form factors \cite{GR01}.

Some interesting opportunities and questions have arisen recently in the
context of singly-Cabibbo-suppressed charm decays.  Through the excellent
photon and charged particle identification capabilities of the CLEO
detector, it has become feasible to study many decays involving $\eta$
and $\eta\,'$ \cite{EVT}.  The FOCUS Collaboration has recently amassed a
large sample of charmed particles produced by high-energy photons at the
Fermilab Tevatron \cite{FOCUS}.  Close and Lipkin \cite{CL02} have recently
identified some puzzles in this sector, including claims for rates for $D^+
\to K^{*+} \ol{K}^0$ \cite{Frab95} and $D^+ \to K^{*+} \ol{K}^{*0}$
\cite{Alb92} of the same order as some Cabibbo-favored two-body modes.
We shall show that it is difficult to understand the first of these.  While
we do not have enough information to analyze charmed particle decays to VV,
we use U-spin to relate these last two rates to those for $D_s^+ \to
\rho^+ \ko$ and $D_s^+ \to \rho^+ K^{*0}$, respectively, which should also
have large rates if the claims are correct and approximate flavor symmetry is
valid.

We recall notation in Section \ref{sec:notn}, and update results of Ref.\ 
\cite{JR99} for Cabibbo-favored decays in Sec.\ \ref{sec:cf}.  We then tabulate
results for Cabibbo-suppressed decays, and discuss specific relations among
these decays (and between them and some Cabibbo-favored processes) in Sec.\ 
\ref{sec:cs}.
We remark briefly on a relation for $VV$ decays in Sec.\ \ref{sec:VV}.  Open
questions are noted in Sec.\ \ref{sec:que}, which concludes.

\section{NOTATION}
\label{sec:notn}

We use the following quark content and phase conventions:
\begin{itemize}
\item{
{\it Charmed mesons}: $D^0=-c\ub$, $D^+=c\db$, $D_s^+=c\sb$;}
\item{
{\it Pseudoscalar mesons}: $\pi^+=u\db$, $\pi^0=(d\db-u\ub)/\sqrt{2}$,
 $\pi^-=-d\ub$, $K^+=u\sb$, $K^0=d\sb$, ${\ol{K}}^0=s\db$, $K^-=-s\ub$,
 $\eta=(s\sb-u\ub-d\db)/\sqrt{3}$,
 $\eta^{\prime}=(u\ub+d\db+2s\sb)/\sqrt{6}$;}
\item{
{\it Vector mesons}: $\rho^+=u\db$, $\rho^0=(d\db-u\ub)/\sqrt{2}$,
 $\rho^-=-d\ub$, $\omega=(u\ub+d\db)/\sqrt{2}$, $K^{*+}=u\sb$,
 $K^{*0}=d\sb$, ${\ol{K}}^{*0}=s\db$, $K^{*-}=-s\ub$, $\phi=s\sb$.}
\end{itemize}

We denote the tree, color-suppressed, exchange, and annihilation amplitudes by
$T$, $C$, $E$, and $A$, respectively.  The exchange and annihilation diagrams
that involve singlet contributions are labeled by $SE$ and $SA$.  Penguin
diagrams should be negligible since $V_{cd}^* V_{ud} \simeq -V_{cs}^* V_{us}$.
For the $PV$ modes, we use the subscripts $P$ and $V$ to refer to those
diagrams with the spectator quark going into a pseudoscalar meson and a vector
meson in the final state, respectively.  To distinguish between Cabibbo-favored
and Cabibbo-suppressed decay modes, the amplitudes associated with the former
are all unprimed, while those with the latter are primed.

The partial width $\Gamma$ for a specific two-body decay to $PP$ is expressed
in terms of an invariant amplitude ${\cal A}$ as
\be
\Gamma(H \to PP) = \frac{p^*}{8 \pi M_H^2}|{\cal A}|^2~,
\ee
where $p^*$ is the center-of-mass (c.m.) 3-momentum of each final particle, and
$M_H$ is the mass of the decaying particle.  The kinematic factor of $p^*$ is
appropriate for the S-wave final state.  The amplitude ${\cal A}$ will thus
have a dimension of (energy).  For $PV$ decays, on the other hand, a P-wave
kinematic factor is appropriate instead, and
\be
\Gamma(H \to PV) = \frac{{p^*}^3}{8 \pi M_H^2}|{\cal A}|^2~.
\ee
In this case, ${\cal A}$ is dimensionless.

In the numerical calculation, we use for the charmed mesons $M_{D^+} = 1.8693
\pm 0.0005$ GeV with $\tau(D^+) = 1051 \pm 13$ fs, $M_{D^0} = 1.8645 \pm
0.0005$ GeV with $\tau(D^0) = 411.7 \pm 2.7$ fs, and $M_{D_s^+} = 1.9685 \pm
0.0006$ GeV with $\tau(D_s^+) = 490 \pm 9$ fs \cite{Hagiwara:pw}.

\section{CABIBBO-FAVORED DECAYS}
\label{sec:cf}

\begin{table}[t]
\caption{Branching ratios and invariant amplitudes for Cabibbo-favored
decays of charmed mesons to two pseudoscalar mesons.
\label{tab:CFPP}}
\begin{center}
\begin{tabular}{c l c c c c}
\hline \hline
Meson & Decay Mode & Representation
     & ${\cal B}$ \cite{Hagiwara:pw} & $p^*$ & $|{\cal A}|$ \\
 & & & ($\%$) & (MeV) & $(10^{-6} {\rm GeV})$ \\
\hline \hline
$D^0$ & $K^- \pi^+$ & $T + E$
     & $3.80 \pm 0.09$ & 861 & $2.48 \pm 0.03$ \\
 & $\ol{K}^0 \pi^0$ & $\frac{1}{\sqrt{2}}(C - E)$
     & $2.28 \pm 0.22$ & 860 & $1.92 \pm 0.09$ \\
 & $\ol{K}^0 \eta$ & $C/\sqrt{3}$
     & $0.76 \pm 0.11$ & 772 & $1.17 \pm 0.08$ \\
 & $\ol{K}^0 \eta\,'$ & $-\frac{1}{\sqrt{6}}(C + 3E)$
     & $1.87 \pm 0.28$ & 565 & $2.15 \pm 0.16$ \\
\hline
$D^+$ & $\ol{K}^0 \pi^+$ & $C + T$
     & $2.77 \pm 0.18$ & 862 & $1.33 \pm 0.04$ \\
\hline
$D_s^+$ & $\ol{K}^0 K^+$ & $C + A$
     & $3.6 \pm 1.1$ & 850 & $2.35 \pm 0.36$ \\
 & $\pi^+ \eta$ & $\frac{1}{\sqrt{3}}(T - 2A)$
     & $1.7 \pm 0.5$ & 902 & $1.57 \pm 0.23$ \\
 & $\pi^+ \eta\,'$ & $\frac{2}{\sqrt{6}}(T + A)$
     & $3.9 \pm 1.0$ & 743 & $2.62 \pm 0.34$ \\
\hline \hline
\end{tabular}
\end{center}
\end{table}
\begin{table}[t]
\caption{Branching ratios and invariant amplitudes for Cabibbo-favored
     decays of charmed mesons to one pseudoscalar and one vector meson.
\label{tab:CFPV}}
\vspace{6pt}
\begin{tabular}{c l c c c c}
\hline \hline
Meson & Decay Mode & Representation
     & ${\cal B}$ \cite{Hagiwara:pw} & $p^*$ & $|{\cal A}|$ \\
 & & & ($\%$) & (MeV) & $(10^{-6})$ \\
\hline \hline
$D^0$ & $K^{*-} \pi^+$ & $T_V + E_P$
     & $6.0 \pm 0.5$ & 711 & $4.83 \pm 0.20$ \\
 & $K^- \rho^+$ & $T_P + E_V$
     & $10.2 \pm 0.8$ & 678 & $6.76 \pm 0.26$ \\
 & $\ol{K}^{*0} \pi^0$ & $\frac{1}{\sqrt{2}}(C_P - E_P)$
     & $2.8 \pm 0.4$ & 709 & $3.31 \pm 0.24$ \\
 & $\ol{K}^0 \rho^0$ & $(C_V - E_V)/\sqrt{2}$
     & $1.47 \pm 0.29$ & 676 & $2.57 \pm 0.25$ \\
 & $\ol{K}^{*0} \eta$ & $\frac{1}{\sqrt{3}}(C_P + E_P - E_V)$
     & $1.8 \pm 0.4$ & 580 & $3.59 \pm 0.40$ \\
 & $\ol{K}^{*0} \eta\,'$ & $-\frac{1}{\sqrt{6}}(C_P + E_P + 2 E_V)$
     & $< 0.10$ &  99 & $< 11.9$ \\
 & $\ol{K}^0 \omega$ & $-\frac{1}{\sqrt{2}}(C_V + E_V)$
     & $2.2 \pm 0.4$ & 670 & $3.20 \pm 0.29$ \\
 & $\ol{K}^0 \phi$ & $-E_P$
     & $0.94 \pm 0.11$ & 520 & $3.05 \pm 0.18$ \\
\hline
$D^+$ & $\ol{K}^{*0} \pi^+$ & $T_V + C_P$
     & $1.92 \pm 0.19$ & 712 & $1.71 \pm 0.08$ \\
 & $\ol{K}^0 \rho^+$ & $T_P + C_V$
     & $6.6 \pm 2.5$ & 680 & $3.40 \pm 0.64$ \\
\hline
$D_s^+$ & $\ol{K}^{*0} K^+$ & $C_P + A_V$
     & $3.3 \pm 0.9$ & 682 & $3.69 \pm 0.50$ \\
 & $\ol{K}^0 K^{*+}$ & $C_V + A_P$
     & $4.3 \pm 1.4$ & 683 & $4.20 \pm 0.68$ \\
 & $\rho^+ \eta$ & $\frac{1}{\sqrt{3}}(T_P - A_P - A_V)$
     & $10.8 \pm 3.1$ & 727 & $6.06 \pm 0.87$ \\
 & $\rho^+ \eta\,'$ & $\frac{1}{\sqrt{6}}(2T_P + A_P + A_V)$
     & $10.1 \pm 2.8$  & 470 & $11.3 \pm 1.6$ \\
 & $\pi^+ \rho^0$ & $\frac{1}{\sqrt{2}}(A_V - A_P)$
     & $< 0.07$ & 827 & $< 0.40$ \\
 & $\pi^+ \omega$ & $\frac{1}{\sqrt{2}}(A_V + A_P)$
     & $0.28 \pm 0.11$ & 822 & $0.81 \pm 0.16$ \\
 & $\pi^+ \phi$ & $T_V$
     & $3.6 \pm 0.9$ & 712 & $3.61 \pm 0.45$ \\
\hline \hline
\end{tabular}
\end{table}

In Tables \ref{tab:CFPP} and \ref{tab:CFPV} we summarize predicted and observed
amplitudes for Cabibbo-favored decays of charmed mesons to $PP$ and $PV$.  The
experimental values are based on Ref.\ \cite{Hagiwara:pw} and supersede those
quoted in Ref.\ \cite{JR99}.  We then extract amplitudes for specific flavor
topologies and their relative phases.  Only the preferred solutions in
Ref.~\cite{JR99} with updated data analysis are quoted in Table
\ref{tab:CFamp}.  These parameters are needed since we will be using flavor
SU(3) to relate them to the singly-Cabibbo-suppressed decays.  For the sign of
relative strong phases, we use the convention that $\delta_{AB}$ means the
angle subtended from the amplitude $B$ to $A$.
\begin{table}[t]
\caption{Preferred solutions of magnitudes and relative phases of the
  invariant amplitudes for the Cabibbo-favored decay modes.
\label{tab:CFamp}}
\vspace{6pt}
\begin{tabular}{ccc}
\hline\hline
Amplitude & Magnitude & Relative Strong Phase \\
$PP$ & ($10^{-6}$ GeV) & \\
\hline
$T$ & $2.67 \pm 0.20$ & --- \\
$C$ & $2.03 \pm 0.15$ & $\delta_{CT} = (-151 \pm 4)^\circ$ \\
$E$ & $1.67 \pm 0.13$ & $\delta_{ET} = (115 \pm 5)^\circ$ \\
$A$ & $1.05 \pm 0.52$ & $\delta_{AT} = (-65 \pm 30)^\circ$ \\
\hline \hline
$PV$ & ($10^{-6}$) & \\
\hline
$T_V$ & $3.61 \pm 0.45$ & --- \\
$C_P$ & $2.44 \pm 0.52$ & $\delta_{C_PT_V} = (-156 \pm 12)^\circ$ \\
$E_P$ & $3.05 \pm 0.18$ & $\delta_{E_PT_V} = (88 \pm 11)^\circ$ \\
\hline
$T_P$ & $6.03 \pm 1.15$ \footnote{Another possible solution is $|T_P| = (4.46
  \pm 1.19) \times 10^{-6}$.  It is disfavored because it gives an even
  unacceptably larger lower bound on $|A\,'_P|$, as will be explained
toward the end of
Sec.\ IV B.}
& --- \\
$C_V$ & $2.74 \pm 0.46$ & $\delta_{C_VT_P} = (-168 \pm 24)^\circ$ \\
$E_V$ & $3.05 \pm 0.18$ & $\delta_{E_VT_P} = (-90 \pm 22)^\circ$ \\
\hline\hline
\end{tabular}
\end{table}
Using the relation $E_V = -E_P$ for the $PV$ modes in Table \ref{tab:CFamp},
one would get the following strong phases for the last three amplitudes, all
relative to $T_V$,
\begin{equation}
\delta_{T_PT_V} = (-3 \pm 25)^\circ ~, \qquad
\delta_{C_VT_V} = (-170 \pm 13)^\circ ~, \qquad
\delta_{E_VT_V} = (-92 \pm 11)^\circ ~.
\end{equation}
It is interesting to observe that in this case $T_P$ and $T_V$ are roughly
pointing in the same direction on the complex plane.  The same is also true for
$C_P$ and $C_V$, but pointing in almost the opposite direction to that of $T_P$
and $T_V$.  $E_P$ and $E_V$, on the other hand, are close to $90^\circ$ from
the above line.

In Ref.\ \cite{JR99} we did not fit amplitudes involving the annihilation
terms $A_P$ and $A_V$.  Normally we would have expected that $A_P = - A_V$
and hence that the decay $D^+ \to \pi^+ \omega$ would be suppressed while
$D^+ \to \pi^+ \rho^0$ would provide information on the magnitude of $A_P$.
This pattern was anticipated some time ago by Lipkin on the basis of a
G-parity argument \cite{HJL89}.  Instead, it is the latter decay which appears
to be suppressed, while the former occurs with a measurable rate.  It may be
that the $\omega$ contains a small admixture of strange quarks, which would
permit it to be produced via a $T_V$ amplitude, or rescattering effects
could induce annihilation-like terms not respecting $A_P = - A_V$.  Other
Cabibbo-favored processes not fitted in this scheme \cite{JR99} include the
decays $D_s^+ \to \rho^+ \eta$ and $D_s^+ \to \rho^+ \eta\,'$.

\section{SINGLY-CABIBBO-SUPPRESSED DECAYS}
\label{sec:cs}

The topological amplitude decomposition of singly-Cabibbo-suppressed two-body
$D$ decays is listed in Table \ref{tab:CSPP} ($PP$ modes) and \ref{tab:CSPV}
($PV$ modes), where the relations $E\,'_V = -E\,'_P$ and $A\,'_V = -A\,'_P$
have been used.
\begin{table}[t]
\caption{Branching ratios and invariant amplitudes for
singly-Cabibbo-suppressed decays of charmed mesons to two pseudoscalar mesons.
\label{tab:CSPP}}
\vspace{6pt}
\begin{tabular}{c l c c c c}
\hline\hline
Meson & Decay Mode & Representation 
     & ${\cal B}$ \cite{Hagiwara:pw} & $p^*$ & $|{\cal A}|$ \\
 & & & ($\times 10^{-3}$) & (MeV) & $(10^{-7} {\rm GeV})$ \\
\hline \hline
$D^0$
  & $\pi^+ \pi^-$ & $-(T\,'+E\,')$
     & $1.43\pm0.07$ & 922 & $4.66 \pm 0.11$ \\
  & $\pi^0 \pi^0$ & $-\frac{1}{\sqrt{2}}(C\,'-E\,')$
     & $0.84\pm0.22$ & 922 & $3.57 \pm 0.47$ \\
  & $K^+ K^-$ & $T\,'+E\,'$
     & $4.12\pm0.14$ & 791 & $8.53 \pm 0.14$ \\
  & $K^0 \ol{K}^0$ & $0$
     & $0.71\pm0.19$ & 788 & $3.55 \pm 0.47$ \\
  & $\pi^0 \eta$ & $\frac{1}{\sqrt{6}}(C\,'-2E\,'-SE\,')$
     & --- & 846 & --- \\
  & $\pi^0 \eta\,'$ & $\frac{1}{\sqrt{3}}(C\,'+E\,'+2SE\,')$
     & --- & 678 & --- \\
  & $\eta \eta$ & $\frac{2\sqrt{2}}{3}(C\,'+SE\,')$
     & --- & 755 & --- \\
  & $\eta \eta\,'$ & $-\frac{1}{3\sqrt{2}}(C\,'+6E\,'+7SE\,')$
     & --- & 537 & --- \\
\hline
$D^+$
  & $\pi^+ \pi^0$ & $-\frac{1}{\sqrt{2}}(T\,'+C\,')$
     & $2.5\pm0.7$ & 925 & $3.86 \pm 0.54$ \\
  & $\pi^+ \eta$ & $\frac{1}{\sqrt{3}}(T\,'+2C\,'+2A\,'+SA\,')$
     & $3.0\pm0.6$ & 848 & $4.41 \pm 0.44$ \\
  & $\pi^+ \eta\,'$ & $-\frac{1}{\sqrt{6}}(T\,'-C\,'+2A\,'+4SA\,')$
     & $5.0\pm1.0$ & 680 & $6.36 \pm 0.64$ \\
  & $K^+ \ol{K}^0$ & $T\,'-A\,'$
     & $5.8\pm0.6$ & 792 & $6.34 \pm 0.33$ \\
\hline
$D_s^+$
  & $\pi^+ K^0$ & $-(T\,'-A\,')$
     & $<8$ & 916 & $<11$ \\
  & $\pi^0 K^+$ & $-\frac{1}{\sqrt{2}}(C\,'+A\,')$
     & --- & 917 & --- \\
  & $\eta K^+$ & $\frac{1}{\sqrt{3}}(T\,'+2C\,'-SA\,')$
     & --- & 835 & --- \\
  & $\eta\,' K^+$ & $\frac{1}{\sqrt{6}}(2T\,'+C\,'+3A\,'+4SA\,')$
     & --- & 646 & --- \\
\hline\hline
\end{tabular}
\vspace{6pt}
\end{table}

\subsection{U-spin relations}

A number of relations between singly-Cabibbo-suppressed amplitudes follow from
the U-spin symmetry interchanging $s$ and $d$ quarks \cite{GR01,MGU}.  The
effective interactions inducing the transitions $c \to s u \bar s$ and $c
\to d u \bar d$ occur with equal and opposite CKM factors, leading to a
term transforming as $U = 1, U_3 = 0$.  One then obtains the following
relations:
\bigskip

\leftline{\it $PP$ decays:}

\be \label{eqn:ppkk}
\ca(D^0 \to \pi^+ \pi^-) = - \ca(D^0 \to K^+ K^-)~,
\ee
\be \label{eqn:kk0}
\ca(D^0 \to \ko \ok) = 0~,
\ee
\be \label{eqn:kkkp}
\ca(D^+ \to K^+ \ok) = - \ca(D_s^+ \to \pi^+ \ko)~.
\ee

\renewcommand{\arraystretch}{0.85}
\begin{table}
\caption{Branching ratios and invariant amplitudes for
  singly-Cabibbo-suppressed decays of charmed mesons to one pseudoscalar and
  one vector meson.
\label{tab:CSPV}}
\begin{center}
\begin{tabular}{c l c c c c}
\hline\hline
Meson & Decay Mode & Representation 
     & ${\cal B}$ \cite{Hagiwara:pw} & $p^*$ & $|{\cal A}|$ \\
 & & & ($\times 10^{-3}$) & (MeV) & $(10^{-7})$ \\
\hline \hline
$D^0$
  & $\pi^+ \rho^-$ & $-(T\,'_V+E\,'_P)$
     & --- & 766 & --- \\
  & $\pi^- \rho^+$ & $-(T\,'_P-E\,'_P)$
     & --- & 766 & --- \\
  & $\pi^0 \rho^0$ & $-\frac12(C\,'_P+C\,'_V)$
     & --- & 767 & --- \\
  & $K^+ K^{*-}$ & $T\,'_V+E\,'_P$
     & $2.0\pm1.1$ & 610 & $11.11 \pm 3.05$ \\
  & $K^- K^{*+}$ & $T\,'_P-E\,'_P$
     & $3.8\pm0.8$ & 610 & $15.31 \pm 1.61$ \\
  & $K^0 \ol{K}^{*0}$ & $-2E\,'_P$
     & $<1.7$ & 605 & $<10.4$ \\
  & $\ol{K}^0 K^{*0}$ & $2E\,'_P$
     & $<0.9$ & 605 & $<7.5$ \\
  & $\pi^0 \omega$ & $\frac12(C\,'_V-C\,'_P+2SE\,'_P)$
     & --- & 761 & --- \\
  & $\pi^0 \phi$ & $\frac{1}{\sqrt{2}}(C\,'_P+SE\,'_P)$
     & $<1.4$ & 644 & $<8.5$ \\
  & $\eta \omega$ & $-\frac{1}{\sqrt{6}}(C\,'_P+2C\,'_V+SE\,'_V+4SE\,'_P)$
     & --- & 648 & --- \\
  & $\eta\,' \omega$ & $\frac{1}{2\sqrt{3}}(C\,'_P-C\,'_V+4SE\,'_V-2SE\,'_P)$
     & --- & 333 & --- \\
  & $\eta \phi$ & $\frac{1}{\sqrt{3}}(C\,'_P-2SE\,'_P+SE\,'_V)$
     & $<2.8$ & 489 & $<18.3$ \\
  & $\eta \rho^0$ & $\frac{1}{\sqrt{6}}(2C\,'_V-C\,'_P-SE\,'_V)$
     & --- & 655 & --- \\
  & $\eta\,' \rho^0$ & $\frac{1}{2\sqrt{3}}(C\,'_V+C\,'_P+4SE\,'_V)$
     & --- & 349 & --- \\
\hline
$D^+$
  & $\pi^+ \rho^0$ & $-\frac{1}{\sqrt{2}}(T\,'_V+C\,'_P-2A\,'_P)$
     & $1.04\pm0.18$ & 769 & $3.55 \pm 0.31$ \\
  & $\pi^0 \rho^+$ & $-\frac{1}{\sqrt{2}}(T\,'_P+C\,'_V+2A\,'_P)$
     & --- & 769 & --- \\
  & $\pi^+ \omega$ & $-\frac{1}{\sqrt{2}}(T\,'_V+C\,'_P+2SA\,'_P)$
     & $<7$ & 763 & $<9.3$ \\
  & $\pi^+ \phi$ & $C\,'_P-SA\,'_P$
     & $6.1\pm0.6$ & 647 & $11.13 \pm 0.55$ \\
  & $\eta \rho^+$ & $\frac{1}{\sqrt{3}}(T\,'_P+2C\,'_V+SA\,'_V)$
     & $<7$ & 659 & $<11.6$ \\
  & $\eta\,' \rho^+$ & $-\frac{1}{\sqrt{6}}(T\,'_P-C\,'_V+4SA\,'_V)$
     & $<5$ & 356 & $<24.7$ \\
  & $K^+ \ol{K}^{*0}$ & $T\,'_V+A\,'_P$
     & $4.2\pm0.5$ & 610 & $10.08 \pm 0.60$ \\
  & $\ol{K}^0 K^{*+}$ & $T\,'_P-A\,'_P$
     & $31\pm14$ & 611 & $27.32 \pm 6.17$ \\
\hline
$D_s^+$
  & $\pi^+ K^{*0}$ & $-(T\,'_V+A\,'_P)$
     & $6.5\pm2.8$ & 773 & $13.57 \pm 2.92$ \\
  & $\pi^0 K^{*+}$ & $-\frac{1}{\sqrt{2}}(C\,'_V-A\,'_P)$
     & --- & 775 & --- \\
  & $K^+ \rho^0$ & $-\frac{1}{\sqrt{2}}(C\,'_P+A\,'_P)$
     & $<2.9$ & 748 & $<9.5$ \\
  & $K^0 \rho^+$ & $-(T\,'_P-A\,'_P)$
     & --- & 746 & --- \\
  & $\eta K^{*+}$ & $\frac{1}{\sqrt{3}}(T\,'_P+2C\,'_V+2A\,'_P-SA\,'_V)$
     & --- & 661 & --- \\
  & $\eta\,' K^{*+}$ & $\frac{1}{\sqrt{6}}(2T\,'_P+C\,'_V+A\,'_P+4SA\,'_V)$
     & --- & 337 & --- \\
  & $K^+ \omega$ & $-\frac{1}{\sqrt{2}}(C\,'_P-A\,'_P-2SA\,'_P)$
     & --- & 741 & --- \\
  & $K^+ \phi$ & $T\,'_V+C\,'_P-A\,'_P+SA\,'_P$
     & $<0.5$ & 607 & $<5.4$ \\
\hline\hline
\end{tabular}
\end{center}
\end{table}

It has been known for some time that the relation (\ref{eqn:ppkk}) fails.
The rate for $D^0 \to K^+ K^-$ is about three times that for $D^0 \to
\pi^+ \pi^-$.  The corresponding amplitudes differ by a factor of about 1.8,
which can be ascribed to the product of a factor $f_K/f_\pi \simeq 1.22$
and a form factor ratio $F_{D \to K}(M_K^2)/F_{D \to \pi}(m_\pi^2) \simeq
1.5$ \cite{GR01,Fact}.  Alternatively, it can be interpreted as saying that the
subprocess $c \to s u \bar s$ leads to lower-multiplicity final states than $c
\to d u \bar d$, since light quarks radiate extra pions easily.  The $c \to s u
\bar s$ subprocess responsible for $D^0 \to K^+ K^-$ has only one light quark
capable of radiating soft pions (in the current-algebra sense), whereas $c \to
d u \bar d$ responsible for $D^0 \to \pi^+ \pi^-$ has three such quarks.  (The
$\bar u$ spectator quark also can radiate soft pions in either case.)
Therefore, one would expect the higher-multiplicity states to be more important
in the fragmentation of the $c \to d u \bar d$ subprocess.  We shall see
presently that an estimate of the amplitude for $D^0 \to \pi^+ \pi^-$ and
$D^0 \to K^+ K^-$ decays based on Cabibbo-favored decays lies {\it between}
the experimental values for these decays.

The amplitude for $D^0 \to \ko \ok$ is predicted to vanish in the U-spin
limit.  Both the initial and final ($J=0$) states have $U=0$, while the
transition operator has $U=1$, as mentioned.  The observed value of this
amplitude is of the same order as the difference between the $D^0 \to \pi^+
\pi^-$ and $D^0 \to K^+ K^-$ amplitudes.  Indeed, if one were to allow
for different effective $c \to s u \bar s$ and $c \to d u \bar d$ transition
strengths {\it in the $E'$ amplitudes alone}, one would obtain the sum rule
\be \label{eqn:dzsum}
\ca(D^0 \to \pi^+ \pi^-) + \ca(D^0 \to K^+ K^-) + \ca(D^0 \to \ko \ok) = 0~,
\ee
which is satisfied when the amplitudes are relatively real with respect to one
another.  However, there is no reason for (\ref{eqn:dzsum}) to hold in general.
The decay $D^0 \to \ko \ok$ simply seems to occur at a level approriate for
SU(3) symmetry breaking in other Cabibbo-suppressed $D^0 \to PP$ decays.

The relation (\ref{eqn:kkkp}) is untested so far.  It predicts a branching
ratio $\b(D_s^+ \to \pi^+ K^0) = (2.8 \pm 0.3) \times 10^{-3}$ on the basis of
$\b(D^+ \to K^+ \ok) = (5.8 \pm 0.6) \times 10^{-3}$ and kinematic correction
factors.  This should be an easy process to observe.  \bigskip

\leftline{\it $PV$ decays:}

\be \label{eqn:prkk1}
\ca(D^0 \to \pi^+ \rho^-) = - \ca(D^0 \to K^+ K^{*-})~,
\ee
\be \label{eqn:prkk2}
\ca(D^0 \to \pi^- \rho^+) = - \ca(D^0 \to K^- K^{*+})~,
\ee
\be \label{eqn:kks}
\ca(D^0 \to \ko \ol{K}^{*0}) = - \ca(D^0 \to \ok K^{*0})~,
\ee
\be \label{eqn:kkspks}
\ca(D^+ \to K^+ \ol{K}^{*0}) = - \ca(D_s^+ \to \pi^+ K^{*0})~,
\ee
\be \label{eqn:kkskr}
\ca(D^+ \to \ok K^{*+}) = - \ca(D_s^+ \to K^0 \rho^+)~.
\ee

The relations (\ref{eqn:prkk1}) and (\ref{eqn:prkk2}) are untested as yet
because of the absence of $D^0 \to \pi^\pm \rho^\mp$ branching ratios.  These
processes should be observable in the CLEO-c detector.  The relation
(\ref{eqn:kks}) should be testable in the presence of an $E\,'_P$ amplitude,
whose magnitude we shall estimate presently.  The relation (\ref{eqn:kkspks})
is satisfied within $1 \sigma$; it predicts $\b(D_s^+ \to \pi^+ K^{*0}) =
(3.6 \pm 0.4) \times 10^{-3}$.  Finally, the relation (\ref{eqn:kkskr}) is
interesting since $\b(D^+ \to \ok K^{*+}) = (3.1 \pm 1.4)\%$ would entail a
predicted branching ratio $\b(D_s^+ \to K^0 \rho^+) = (2.4 \pm 1.1)\%$.  We
shall see, however, that it is difficult to understand the large branching
ratio for $D^+ \to \ok K^{*+}$ when extrapolating from the Cabibbo-favored
$PV$ decays of charmed particles \cite{CL02}, even when one allows for the
most favorable possible interference between contributing amplitudes.

\subsection{Relations between Cabibbo-favored and singly-Cabibbo-suppressed
decays}

We now make use of the amplitudes determined in Ref.\ \cite{JR99} and updated
in Sec.\ \ref{sec:cf} for Cabibbo-favored $PP$ and $PV$ decays to predict the
magnitudes and phases of amplitudes for singly-Cabibbo-suppressed processes.
We shall see that with the single exception of $D^+ \to \ok K^{*+}$, all
results are consistent with a flavor SU(3) symmetry whose breaking does not
exceed expected limits.  The magnitudes of the topological amplitudes for
singly-Cabibbo-suppressed modes can be obtained from those for Cabibbo-favored
ones listed in Table \ref{tab:CFamp} by multiplying a Cabibbo suppression
factor of $\lambda \simeq 0.2256$.  We assume the relative strong phases stay
the same.  The resulting amplitudes are shown in Table \ref{tab:CSamp}.

\renewcommand{\arraystretch}{1.0}
\begin{table}[t]
\caption{Real and imaginary parts of the invariant amplitudes for the
singly-Cabibbo-suppressed decay modes. It is assumed that $T'$ and $T'_V$
are purely real.
\label{tab:CSamp}}
\vspace{6pt}
\begin{tabular}{c c c c c c}
\hline \hline
Amplitude & Re & Im & Amplitude & Re & Im \\
$PP$ & $(10^{-7} {\rm GeV})$ & $(10^{-7} {\rm GeV})$ & $PV$ & $(10^{-7})$
& $(10^{-7})$ \\
\hline
$T'$ & 6.02 & 0 & $T'_V$ & 8.14 & 0 \\
$C'$ & $-4.01$ & $-2.22$ & $C'_P$ & $-5.03$ & $-2.24$ \\
$E'$ & $-1.59$ & 3.41 & $E'_P$ & 0.30 & 6.88 \\
$A'$ & 1.00 & $-2.15$ & $T'_P$ & 13.6 & $-0.60$ \\
& & & $C'_V$ & $-6.09$ & $-1.07$ \\
& & & $E'_V$ & $-0.30$ & $-6.88$ \\
\hline \hline
\end{tabular}
\end{table}

In the singly-Cabibbo-suppressed $PP$ decays, some modes can be directly
related to their counterparts in the Cabibbo-favored ones.  Assuming $SU(3)$
symmetry, we obtain
\begin{eqnarray*}
&& |{\cal A}(D^0 \to \pi^+ \pi^-)| = |{\cal A}(D^0 \to K^+ K^-)|
= \lambda |{\cal A}(D^0 \to K^- \pi^+)|
\simeq (5.60 \pm 0.07) \times 10^{-7} {\rm GeV}~, \\
&& |{\cal A}(D^0 \to \pi^0 \pi^0)|
= \lambda |{\cal A}(D^0 \to \ol{K}^0 \pi^0)|
\simeq (4.34 \pm 0.21) \times 10^{-7} {\rm GeV}~, \\
&& |{\cal A}(D^+ \to \pi^+ \pi^0)|
= \frac{\lambda}{\sqrt{2}} |{\cal A}(D^+ \to \ol{K}^0 \pi^+)|
\simeq (2.12 \pm 0.07) \times 10^{-7} {\rm GeV}~, \\
&& |{\cal A}(D_s^+ \to \pi^0 K^+)|
= \frac{\lambda}{\sqrt{2}} |{\cal A}(D_s^+ \to \ol{K}^0 K^+)|
\simeq (3.76 \pm 0.57) \times 10^{-7} {\rm GeV}~.
\end{eqnarray*}
Both $D^+ \to K^+ \ol{K}^0$ and $D_s^+ \to \pi^+ K^0$ involve the combination
$T\,'-A\,'$, which does not have a counterpart in the Cabibbo-favored modes.
Therefore, we use the values given in Table \ref{tab:CSamp} to estimate the
magnitude of their amplitude:
\begin{eqnarray*}
|{\cal A}(D^+ \to K^+ \ol{K}^0)| = |{\cal A}(D_s^+ \to \pi^+ K^0)|
\simeq (5.47 \pm 1.30) \times 10^{-7} {\rm GeV}~.
\end{eqnarray*}

In the $PV$ decays, the following results are obtained:
\begin{eqnarray*}
&& |{\cal A}(D^0 \to \pi^+ \rho^-)| = |{\cal A}(D^0 \to K^+ K^{*-})|
= \lambda |{\cal A}(D^0 \to K^{*-} \pi^+)|
\simeq (10.90 \pm 0.45) \times 10^{-7}~, \\
&& |{\cal A}(D^0 \to \pi^- \rho^+)| = |{\cal A}(D^0 \to K^- K^{*+})|
= \lambda |{\cal A}(D^0 \to K^- \rho^+)|
\simeq (15.25 \pm 0.60) \times 10^{-7}~, \\
&& |{\cal A}(D^0 \to K^0 \ol{K}^{*0})| = |{\cal A}(D^0 \to \ol{K}^0 K^{*0})|
= 2 \lambda |{\cal A}(D^0 \to \ol{K}^0 \phi)|
\simeq (13.78 \pm 0.81) \times 10^{-7}~, \\
&& |{\cal A}(D^0 \to \pi^0 \rho^0)|
\simeq (5.83 \pm 0.78) \times 10^{-7}~,
\end{eqnarray*}
where the last line is computed directly using Table \ref{tab:CSamp}.  It is
seen that all the above predicted amplitude magnitudes agree well with those
inferred from the measured branching ratios, apart from small differences
that can be attributed to $SU(3)$ breaking.  Table \ref{tab:CMPPE} summarizes
the comparison of predicted and experimental amplitudes.

\begin{table}
\caption{Comparisons between predicted amplitudes based on
Cabibbo-favored decays and the experimental values for
singly-Cabibbo-suppressed decays of charmed mesons.
\label{tab:CMPPE}}
\vspace{6pt}
\begin{tabular}{c c c c}
\hline \hline
Meson & Decay Mode & Prediction & Experimental value \\
& $PP$ & $(10^{-7} {\rm GeV})$ & $(10^{-7} {\rm GeV})$ \\
\hline
$D^0$ & $\pi^+\pi^-$ & $5.60 \pm 0.07$ & $4.66 \pm 0.11$ \\
& $\pi^0\pi^0$ & $4.34 \pm 0.21$ & $3.57 \pm 0.47$ \\
& $K^+K^-$ & $5.60 \pm 0.07$ & $8.53 \pm 0.14$ \\
& $K^0 \overline{K}^0$ & 0 & $3.55 \pm 0.47$ \\
\hline
$D^+$ & $\pi^+\pi^0$ & $2.12 \pm 0.07$ & $3.86 \pm 0.54$ \\
 & $K^+ \overline{K}^0$ & $5.47 \pm 1.30$ & $6.34 \pm 0.33$ \\
\hline
$D_s^+$ & $\pi^+ K^0$ & $5.47 \pm 1.30$ & $<11$ \\
\hline \hline
& $PV$ & $(10^{-7})$ & $(10^{-7})$ \\
\hline
$D^0$ & $K^+K^{*-}$ & $10.90 \pm 0.45$ & $11.11 \pm 3.05$ \\
& $K^-K^{*+}$ & $15.25 \pm 0.60$ & $15.31 \pm 1.61$ \\
& $K^0\overline{K}^{*0}$ & $13.78 \pm 0.81$ & $<10.4$ \\
& $\overline{K}^0K^{*0}$ & $13.78 \pm 0.81$ & $<7.5$ \\
\hline \hline
\end{tabular}
\end{table}

An upper bound on $A_P$ can be extracted from Cabibbo-favored modes: $|A_P|
\le (|{\cal A}(D_s^+ \to \pi^+ \rho^0)|^2 + |{\cal A}(D_s^+ \to \pi^+
\omega)|^2)^{1/2} \le 1.4 \times 10^{-6}$ at the $3 \sigma$ level.  This in
turn implies that the corresponding singly-Cabibbo-suppressed amplitude
satisfies
\be
\label{eq:APp}
|A\,'_P| = \lambda |A_P| \le 3.1 \times 10^{-7}~.
\ee
Other information about the contribution of $A\,'_P$ can be directly learned
from, for example, the decay mode $D^+ \to \pi^+ \rho^0$.  In order to
reproduce the $1 \sigma$ lower limit on its amplitude, using
\begin{eqnarray*}
|{\cal A}(D^+ \to \pi^+ \rho^0)|
= |\sqrt{2} A\,'_P - \frac{\lambda}{\sqrt{2}}
                     {\cal A}(D^+ \to \ol{K}^{*0} \pi^+)|~,
\end{eqnarray*}
one must have $|A\,'_P| \ge 0.36 \times 10^{-7}$, assuming that the two
contributions interfere constructively.  Since $|T\,'_V| \simeq (8.2 \pm 2.6)
\times 10^{-7}$, a small $|A\,'_P|$ of about $2 \times 10^{-7}$
improves agreement with the extracted amplitude of $D^+ \to K^+ \ol{K}^{*0}$,
assuming constructive interference. However, there is trouble when one tries to
interpret the experimental data for $D^+ \to \ol{K}^0 K^{*+}$.  With $|T\,'_P|
= \lambda |T_P| \simeq (13.6 \pm 2.6) \times 10^{-7}$, one would need
$|A\,'_P| \agt (13.7 \pm 6.7) \times 10^{-7}$ in order to reach the
experimental result in Table \ref{tab:CSPV}, where the lower bound on $A\,'_P$
assumes maximal constructive interference.  This apparently contradicts
the upper bound (\ref{eq:APp}) obtained from the Cabibbo-favored modes.  Since
currently the branching ratio of $D^+ \to \ol{K}^0 K^{*+}$ is measured only at
a level of slightly more than $2\sigma$ \cite{Frab95}, a definite conclusion
cannot be drawn without more statistics.

\subsection{Triangle and quadrangle relations}

From the $PP$ modes listed in Table \ref{tab:CSPP}, one can find the following
sum rules:
\begin{eqnarray}
&& \sqrt{2}\ca(D^+ \to \pi^+ \pi^0) - \sqrt{2}\ca(D^0 \to \pi^0 \pi^0)
- \ca(D^0 \to \pi^+ \pi^-) = 0 ~, \nonumber \\
&& \sqrt{2}\ca(D^+ \to \pi^+ \pi^0) + \ca(D^+ \to K^+ \ol{K}^0)
- \sqrt{2}\ca(D_s^+ \to \pi^0 K^+) = 0 ~, \nonumber \\
&& \sqrt{2}\ca(D^+ \to \pi^+ \pi^0) - \ca(D_s^+ \to \pi^+ K^0)
- \sqrt{2}\ca(D_s^+ \to \pi^0 K^+) = 0 ~, \nonumber \\
&& 2\sqrt{2}\ca(D^+ \to \pi^+ \eta) + \ca(D^+ \to \pi^+ \eta\,')
+ \sqrt{6}\ca(D^+ \to K^+ \ol{K}^0) + 3\sqrt{3}\ca(D^+ \to \pi^+ \pi^0) = 0 ~,
\nonumber \\
&& 2\sqrt{2}\ca(D_s^+ \to \eta K^+) + \ca(D_s^+ \to \eta\,' K^+)
+ \sqrt{6}\ca(D_s^+ \to \pi^+ K^0) + 3\sqrt{3}\ca(D_s^+ \to \pi^0 K^+) = 0 ~.
\nonumber
\end{eqnarray}
Moreover, any three amplitudes selected from $\ca(D^0 \to \pi^0 \pi^0)$,
$\ca(D^0 \to \pi^0 \eta)$, $\ca(D^0 \to \pi^0 \eta\,')$, $\ca(D^0 \to \eta
\eta)$, and $\ca(D^0 \to \eta \eta\,')$ with appropriate coefficients can form
a triangle.

\subsection{Decays involving $\eta$ and $\eta\,'$}

The amplitudes for decays involving $\eta$ and $\eta\,'$ contain unknown
contributions corresponding to disconnected quark diagrams, such as $SE\,'$
and $SA\,'$ in the decays to $PP$.  A satisfactory description of
Cabibbo-favored decays to $PP$ was obtained in Ref.\ \cite{JR99} without
the help of such contributions, but the Cabibbo-favored decays to $PV$
final states involving such contributions were not seen to follow a
pattern describable through the flavor-topology approach.  In the present
subsection we discuss a test for the amplitudes $SE\,'$ and $SA\,'$ which can
determine whether a flavor-topology description is suitable for
singly-Cabibbo-suppressed decays of charmed mesons to $PP$.

\begin{figure}[t]
\centerline{\includegraphics[height=4in]{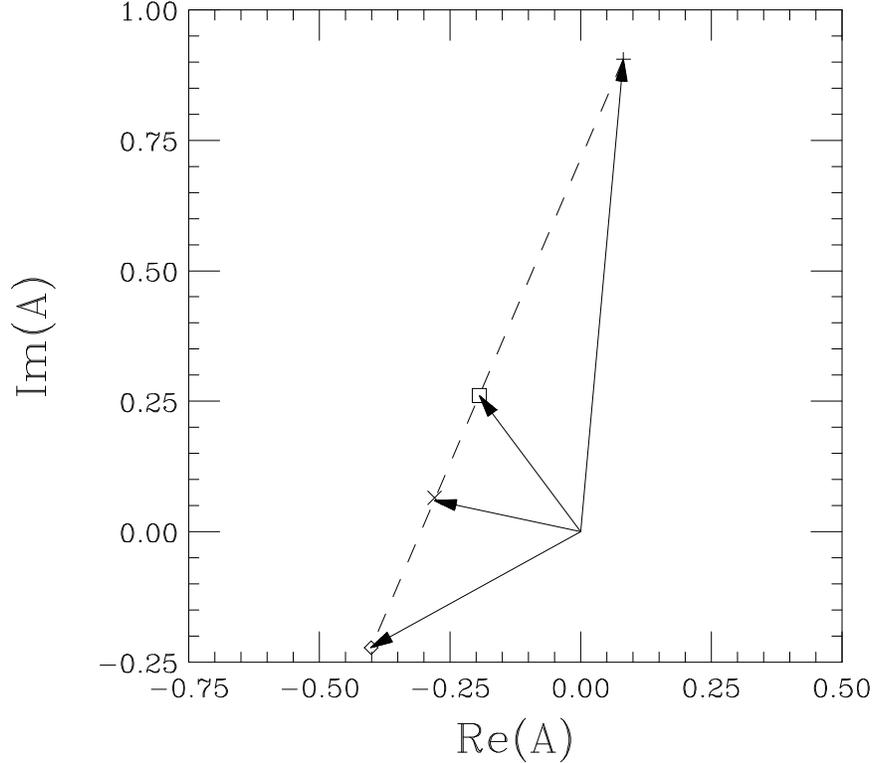}}
\caption{Real and imaginary parts of amplitudes for $D^0$ decays to
$PP$ final states involving $\eta$ and/or $\eta\,'$.  The origin may be shifted
by an arbitrary amount $SE\,'$.  $+$:  $-\sx \ca(D^0 \to \pi^0 \eta)$;
$\times$:  $(\st/2) \ca(D^0 \to \pi^0 \eta\,')$; $\diamond$: $(3/2 \s) \ca(D^0
\to \eta \eta)$; $\Box$: $-(3 \s/7) \ca(D^0 \to \eta \eta\,')$.
\label{fig:d0pp}}
\end{figure}

\begin{figure}[t]
\centerline{\includegraphics[height=4in]{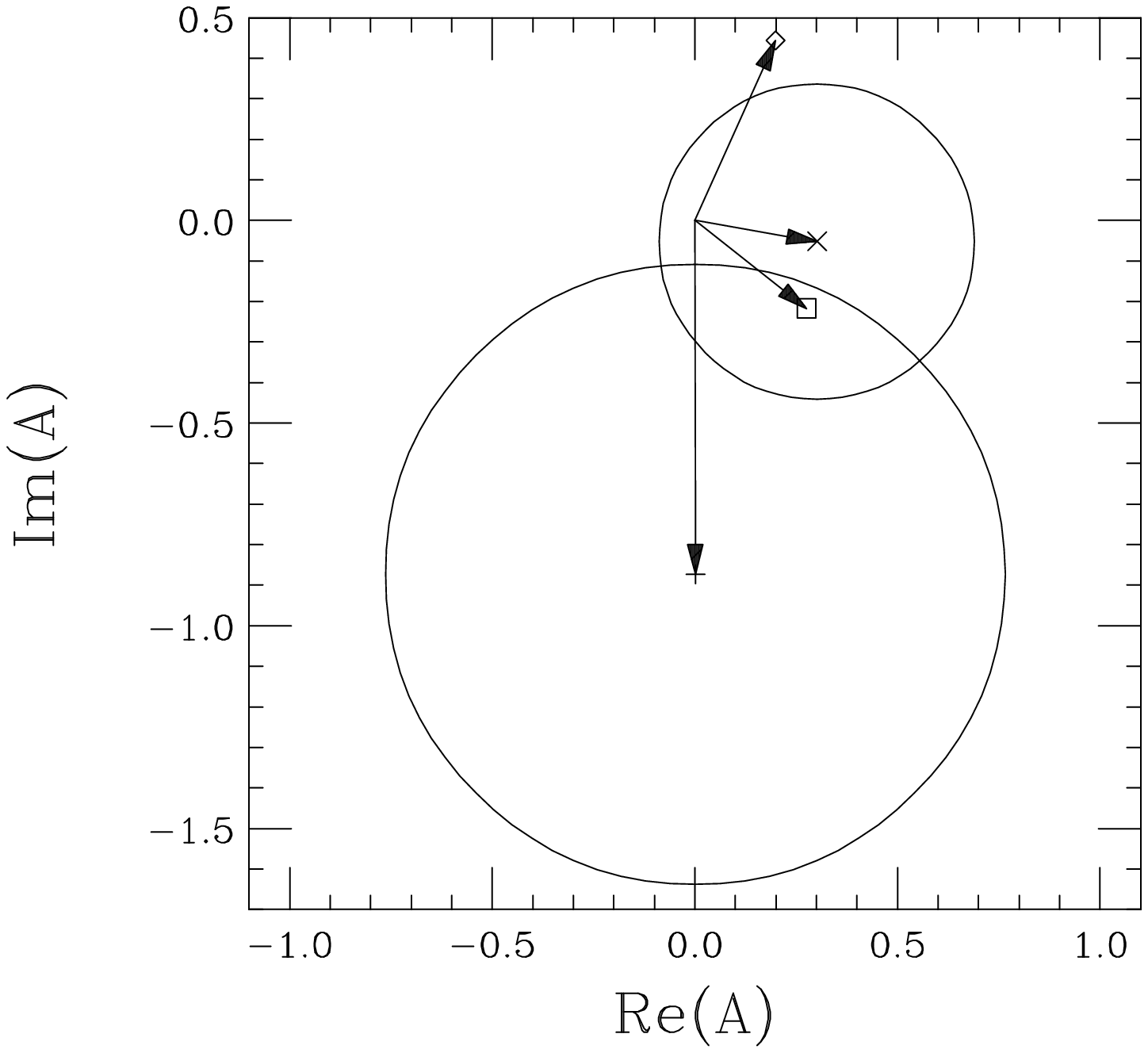}}
\caption{Real and imaginary parts of amplitudes for $D^+$ and $D_s^+$ decays to
$PP$ final states involving $\eta$ and/or $\eta\,'$.  The origin may be shifted
by an arbitrary amount $SA\,'$.  $+$:  $\st \ca(D^+ \to \pi^+ \eta)$; $\times$:
$-(\sx/4) \ca(D^+ \to \pi^+ \eta\,')$; $\diamond$: $-\st \ca(D_s^+ \to
\eta K^+)$; $\Box$: $(\sx/4) \ca(D_s^+ \to \eta\,' K^+)$.  Circles about the
points for $D^+ \to \pi^+ \eta$ and $D^+ \to \pi^+ \eta\,'$ denote central
values of the corresponding magnitudes.
\label{fig:dcpp}}
\end{figure}

We express all amplitudes involving $\eta$ or $\eta\,'$ in Table IV in terms of
an unknown parameter $SE\,'$ or $SA\,'$ with unit coefficient:  For example,
\be
-\sx \ca(D^0 \to \pi^0 \eta\,') = 2E\,' - C\,' + SE\,'~~,
\ee
\be
\frac{\st}{2} \ca(D^0 \to \pi^0 \eta\,') = \frac12 (C\,' + E\,') + SE\,'~~,
\ee
\be
\frac{3}{2 \s} \ca(D^0 \to \eta \eta) = C\,' + SE\,'~~,
\ee
\be
-\frac{3 \s}{7} \ca(D^0 \to \eta \eta\,') = \frac17 (C\,' + 6E\,') + SE\,'~~,
\ee
with four similar expressions (two for $D^+$ and two for $D_s^+$) involving
$SA\,'$.  Assuming $SE\,' = SA\,' = 0$ one can then plot these expressions in
the complex plane, obtaining figures whose origins can be shifted by an
amount corresponding to the unknown amplitude $SE\,'$ or $SA\,'$.  The amplitudes
plotted are summarized in Table \ref{tab:etas} and described in Figs.\
\ref{fig:d0pp} and \ref{fig:dcpp}.

\begin{table}[t]
\caption{Complex amplitudes describing singly-Cabibbo-suppressed charmed
meson decays to $PP$ involving $\eta$ and/or $\eta\,'$.  Real and imaginary
parts of amplitudes are given in units of $10^{-6}$ GeV.  An additional
unknown term $SE\,'$ contributes to each of the first four decays and $SA\,'$
to the last four.
\label{tab:etas}}
\begin{center}
\begin{tabular}{c c r r} \hline \hline
Amplitude   & Expression & Re & Im \\ \hline
$-\sx \ca(D^0 \to \pi^0 \eta)$ & $2E\,'-C\,'$
     & 0.082 & 0.905 \\ 
$-\frac{\st}{2} \ca(D^0 \to \pi^0 \eta\,')$ & $\frac12(C\,' + E\,')$
     & $-0.280$ & 0.060 \\
$\frac{3}{2 \s} \ca(D^0 \to \eta \eta)$ & $C\,'$
     & $-0.401$ & $-0.222$ \\
$-\frac{3 \s}{7} \ca(D^0 \to \eta \eta\,')$ & $\frac17(C\,' + 6E\,')$
     & $-0.194$ & 0.261 \\
$\st \ca(D^+ \to \pi^+ \eta)$ & $T\,'+2C\,'+2A\,'$
     & $0.001$ & $-0.873$ \\
$-\frac{\sx}{4} \ca(D^+ \to \pi^+ \eta\,')$ & $\frac14(T\,'-C\,'+2A\,')$
     & 0.301 & $-0.052$ \\
$-\st \ca(D_s^+ \to \eta K^+)$ & $-(T\,'+2C\,')$
     & 0.199 & 0.444 \\
$\frac{\sx}{4} \ca(D_s^+ \to \eta\,' K^+)$ & $\frac14(2T\,'+C\,'+3A\,')$
     & 0.276 & $-0.217$ \\
\hline \hline
\end{tabular}
\end{center}
\end{table}

The points for these $D^0$ decays all lie on a line, since the coefficients of
$C\,'$ and $E\,'$ always sum to 1.  This is another way of expressing the
linear dependence of the various decays mentioned in the previous subsection.
In the case of $D^+$ and $D_s^+$ decays this linear dependence is not present.

The rates for $D^+ \to \pi^+ \eta$ and $D^+ \to \pi^+ \eta\,'$ have been
measured.  Consequently, one may use them to draw circles about the
corresponding points to search for common intersections.  The line between
each of these common intersection points and the origin corresponds to the
complex amplitude $SA\,'$ needed to reproduce the data.  One solution
corresponds
to very small $SA\,'$, while the other corresponds to a value comparable to the
other amplitudes.  Measurement of rates for $D_s^+ \to \eta K^+$ and $D_s^+
\to \eta\,' K^+$ will permit a choice between these two solutions and a test of
consistency of the description.  A corresponding construction also will clearly 
be possible for $D^0$ decays once these are measured.

In principle similar techniques would be suitable for $PV$ decays.  The decays
$D^0 \to \pi^0 \omega$ and $D^0 \to \pi^0 \phi$ involve just the one unknown
singlet amplitude ${SE}'_P$, allowing a two-fold solution in the manner
of Fig.\ \ref{fig:dcpp}.  Similarly, the decays $D^0 \to \eta \rho^0$ and
$D^0 \to \eta\,' \rho^0$ involve ${SE}'_V$, and again there will be a two-fold
solution.  One can then test whether the four possible combinations of these
solutions are compatible with the observed branching ratios for
$D^0 \to \eta \omega$, $D^0 \to \eta\,' \omega$, and $D^0 \to \eta \phi$, which
involve both ${SE}'_P$ and ${SE}'_V$.

In the case of $D^+$ and $D_s^+$ singly-Cabibbo-suppressed decays to $PV$, the
presence of $A\,'_P$ and $A\,'_V$ in many amplitudes makes a similar program
problematic.  Without information on these quantities, which we found
difficult to extract from Cabibbo-favored decays, the best one can do is
to extract two possible solutions for ${SA}'_P$ from the decay rates for
$D^+ \to \pi^+ \omega$ and $D^+ \to \pi^+ \phi$, and two possible solutions
for ${SA}'_V$ from the decay rates for $D^+ \to \eta \rho^+$ and $D^+ \to
\eta\,' \rho^+$.   One can, at least, see whether there is a need for
disconnected diagrams in these processes.

\section{REMARKS ON $VV$ DECAYS}
\label{sec:VV}

The branching ratio for $D^+ \to K^{*+} \ol{K}^{*0}$ appears to be enhanced
beyond that for a typical singly-Cabibbo-suppressed decay \cite{CL02}.  The
amplitude relation
\be \label{eqn:VVU}
\ca(D^+ \to K^{*+} \ol{K}^{*0}) = - \ca(D_s^+ \to \rho^+ K^{*0})
\ee
should hold separately for each partial wave ($L=0,~1,~2$) as a consequence of
U-spin.  The kinematic correction factors behave as ${(p^*)}^{2L+1}/M_H^2$,
so they cannot be directly applied unless we know the partial-wave
decomposition of the decays.  However, we can obtain a lower limit on the
predicted branching ratio for $D_s^+ \to \rho^+ K^{*0}$ by assuming that
the decays are dominated by $L=0$.  Given that $p^* = 273$ MeV for the $D^+$
decay and 524 MeV for the $D_s^+$ decay, we find that $\b(D^+ \to K^{*+}
\ol{K}^{*0}) = (2.6 \pm 1.1)\%$ implies $\b(D_s^+ \to \rho^+ K^{*0}) = (2.1
\pm 0.9)\%$.  If components with $L \ne 0$ are present this value
becomes a lower bound.

\section{OPEN QUESTIONS AND SUMMARY}
\label{sec:que}

We have some evidence that the flavor-topology approach is limited in
usefulness from the failure of the Cabibbo-favored decays $D_s^+ \to \pi^+
\rho^0$ and $D_s^+ \to \pi^+ \omega$ to fit any reasonble pattern for the
amplitudes $A_P$ and $A_V$.  Furthermore, it appears that disconnected quark
diagrams, neglected in Ref.\ \cite{JR99},  appear necessary to fit the
large branching ratios claimed for $D_s^+ \to \pi^+ \eta$ and $D_s^+ \to \pi^+
\eta\,'$.  One would expect analogues of these puzzles to appear in the
singly-Cabibbo-suppressed decays.  Certainly the processes $D^+ \to K^{*+}
\ok$ and $D^+ \to K^{*+} \ol{K}^{*0}$ noted by Close and Lipkin \cite{CL02}
are the most prominent candidates for such puzzles.  It will be interesting
to see the progress of future experimental studies, for example at CLEO-c,
of these decays.

We have shown that aside from the two decays just noted, a reasonable
description of $PP$ and $PV$ singly-Cabibbo-suppressed decays of charmed
mesons appears possible by extrapolation from the Cabibbo-favored decays.
As in the case of Cabibbo-favored decays, various amplitudes have non-trivial
relative strong phases, indicating that these amplitudes are probably generated
by final-state interactions governed by long-distance physics.

Decays involving $\eta$ and $\eta\,'$ can be described if one is prepared to
consider flavor topologies involving disconnected diagrams.  The magnitudes
of such amplitudes remain to be studied, but there are enough processes
that once a few of them have beem measured, predictions will be possible for
the remaining ones.  Such studies bear the promise of useful insights on the
strong interactions governing final-state interactions in charm decays, and
may also shed indirect light on such interactions at the higher energies
characterizing the decays of hadrons containing $b$ quarks.

\section*{ACKNOWLEDGMENTS}

We thank H. J. Lipkin for discussions. J. L. R. is grateful to the Theory Group
at Argonne National Laboratory for hospitality during part of this study.
This work was supported in part by the United States Department of Energy, High
Energy Physics Division, through Grant No.\ DE-FG02-90ER-40560 and Contract
No.\ W-31-109-ENG-38.

\def \ajp#1#2#3{Am.~J.~Phys.~{\bf#1}, #2 (#3)}
\def \apny#1#2#3{Ann.~Phys.~(N.Y.) {\bf#1}, #2 (#3)}
\def \app#1#2#3{Acta Phys.~Polonica {\bf#1}, #2 (#3)}
\def \arnps#1#2#3{Ann.~Rev.~Nucl.~Part.~Sci.~{\bf#1}, #2 (#3)}
\def \art{and references therein}
\def \b97{{\it Beauty '97}, Proceedings of the Fifth International
Workshop on $B$-Physics at Hadron Machines, Los Angeles, October 13--17,
1997, edited by P. Schlein}
\def \carg{{\it Masses of Fundamental Particles -- Carg\`ese 1996}, edited by
M. L\'evy \ite, NATO ASI Series B:  Physics Vol.~363 (Plenum, New York, 1997)}
\def \cmp#1#2#3{Commun.~Math.~Phys.~{\bf#1}, #2 (#3)}
\def \cmts#1#2#3{Comments on Nucl.~Part.~Phys.~{\bf#1}, #2 (#3)}
\def \corn93{{\it Lepton and Photon Interactions:  XVI International
Symposium, Ithaca, NY August 1993}, AIP Conference Proceedings No.~302,
ed.~by P. Drell and D. Rubin (AIP, New York, 1994)}
\def \cn{Collaboration}
\def \cp89{{\it CP Violation,} edited by C. Jarlskog (World Scientific,
Singapore, 1989)}
\def \dpff{{\it The Fermilab Meeting -- DPF 92} (7th Meeting of the
American Physical Society Division of Particles and Fields), 10--14
November 1992, ed. by C. H. Albright \ite~(World Scientific, Singapore,
1993)}
\def \dpf94{DPF 94 Meeting, Albuquerque, NM, Aug.~2--6, 1994}
\def \efi{Enrico Fermi Institute Report No. EFI}
\def \el#1#2#3{Europhys.~Lett.~{\bf#1}, #2 (#3)}
\def \epjc#1#2#3{Eur.~Phys.~J.~C {\bf#1}, #2 (#3)}
\def \f79{{\it Proceedings of the 1979 International Symposium on Lepton
and Photon Interactions at High Energies,} Fermilab, August 23-29, 1979,
ed.~by T. B. W. Kirk and H. D. I. Abarbanel (Fermi National Accelerator
Laboratory, Batavia, IL, 1979}
\def \hb87{{\it Proceeding of the 1987 International Symposium on Lepton
and Photon Interactions at High Energies,} Hamburg, 1987, ed.~by W. Bartel
and R. R\"uckl (Nucl. Phys. B, Proc. Suppl., vol. 3) (North-Holland,
Amsterdam, 1988)}
\def \ib{{\it ibid.}~}
\def \ibj#1#2#3{{\it ibid.}~{\bf#1}, #2 (#3)}
\def \ichep72{{\it Proceedings of the XVI International Conference on High
Energy Physics}, Chicago and Batavia, Illinois, Sept. 6--13, 1972,
edited by J. D. Jackson, A. Roberts, and R. Donaldson (Fermilab, Batavia,
IL, 1972)}
\def \ijmpa#1#2#3{Int.~J.~Mod.~Phys.~A {\bf#1}, #2 (#3)}
\def \ite{{\it et al.}}
\def \jmp#1#2#3{J.~Math.~Phys.~{\bf#1}, #2 (#3)}
\def \jpg#1#2#3{J.~Phys.~G {\bf#1}, #2 (#3)}
\def \lkl87{{\it Selected Topics in Electroweak Interactions} (Proceedings
of the Second Lake Louise Institute on New Frontiers in Particle Physics,
15--21 February, 1987), edited by J. M. Cameron \ite~(World Scientific,
Singapore, 1987)}
\def \KEK#1{{\it Flavor Physics} (Proceedings of the Fourth International
Conference on Flavor Physics, KEK, Tsukuba, Japan, 29--31 October 1996),
edited by Y. Kuno and M. M. Nojiri, Nucl.~Phys.~B Proc.~Suppl.~{\bf 59},
#1 (1997)}
\def \ky85{{\it Proceedings of the International Symposium on Lepton and
Photon Interactions at High Energy,} Kyoto, Aug.~19-24, 1985, edited by M.
Konuma and K. Takahashi (Kyoto Univ., Kyoto, 1985)}
\def \mpla#1#2#3{Mod.~Phys.~Lett.~A {\bf#1}, #2 (#3)}
\def \nc#1#2#3{Nuovo Cim.~{\bf#1}, #2 (#3)}
\def \nima#1#2#3{Nucl.~Instr.~Meth.~A {\bf#1}, #2 (#3)}
\def \np#1#2#3{Nucl.~Phys.~{\bf#1}, #2 (#3)}
\def \npbps#1#2#3{Nucl.~Phys.~B (Proc.~Suppl.) {\bf#1}, #2 (#3)}
\def \pc{private communication}
\def \pisma#1#2#3#4{Pis'ma Zh.~Eksp.~Teor.~Fiz.~{\bf#1}, #2 (#3) [JETP
Lett. {\bf#1}, #4 (#3)]}
\def \pl#1#2#3{Phys.~Lett.~{\bf#1}, #2 (#3)}
\def \plb#1#2#3{Phys.~Lett.~B {\bf#1}, #2 (#3)}
\def \pr#1#2#3{Phys.~Rev.~{\bf#1}, #2 (#3)}
\def \pra#1#2#3{Phys.~Rev.~A {\bf#1}, #2 (#3)}
\def \prd#1#2#3{Phys.~Rev.~D {\bf#1}, #2 (#3)}
\def \prl#1#2#3{Phys.~Rev.~Lett.~{\bf#1}, #2 (#3)}
\def \prp#1#2#3{Phys.~Rep.~{\bf#1}, #2 (#3)}
\def \ptp#1#2#3{Prog.~Theor.~Phys.~{\bf#1}, #2 (#3)}
\def \rmp#1#2#3{Rev.~Mod.~Phys.~{\bf#1}, #2 (#3)}
\def \rp#1{~~~~~\ldots\ldots{\rm rp~}{#1}~~~~~}
\def \si90{25th International Conference on High Energy Physics, Singapore,
Aug. 2-8, 1990}
\def \slc87{{\it Proceedings of the Salt Lake City Meeting} (Division of
Particles and Fields, American Physical Society, Salt Lake City, Utah,
1987), ed.~by C. DeTar and J. S. Ball (World Scientific, Singapore, 1987)}
\def \slac89{{\it Proceedings of the XIVth International Symposium on
Lepton and Photon Interactions,} Stanford, California, 1989, edited by M.
Riordan (World Scientific, Singapore, 1990)}
\def \smass82{{\it Proceedings of the 1982 DPF Summer Study on Elementary
Particle Physics and Future Facilities}, Snowmass, Colorado, edited by R.
Donaldson, R. Gustafson, and F. Paige (World Scientific, Singapore, 1982)}
\def \smass90{{\it Research Directions for the Decade} (Proceedings of the
1990 Summer Study on High Energy Physics, June 25 -- July 13, Snowmass,
Colorado), edited by E. L. Berger (World Scientific, Singapore, 1992)}
\def \stone{{\it B Decays}, edited by S. Stone (World Scientific,
Singapore, 1994)}
\def \tasi90{{\it Testing the Standard Model} (Proceedings of the 1990
Theoretical Advanced Study Institute in Elementary Particle Physics,
Boulder, Colorado, 3--27 June, 1990), edited by M. Cveti\v{c} and P.
Langacker (World Scientific, Singapore, 1991)}
\def \vanc{29th International Conference on High Energy Physics, Vancouver,
23--31 July 1998}
\def \yaf#1#2#3#4{Yad.~Fiz.~{\bf#1}, #2 (#3) [Sov.~J.~Nucl.~Phys.~{\bf #1},
#4 (#3)]}
\def \zhetf#1#2#3#4#5#6{Zh.~Eksp.~Teor.~Fiz.~{\bf #1}, #2 (#3) [Sov.~Phys.
-- JETP {\bf #4}, #5 (#6)]}
\def \zpc#1#2#3{Zeit.~Phys.~C {\bf#1}, #2 (#3)}

\end{document}